\documentclass[fleqn,twoside,twocolumn,nofootinbib,showkeys]{revtex4} 
\usepackage[sec,nocpr]{ujp} 

\usepackage{xspace}
\usepackage{url}

\usepackage{color,units}

\newcommand{\eq}[1]{\begin{equation} #1 \end{equation}}
\newcommand{\nfw}{{\textsc{nfw}}}
\newcommand{\tnfw}{{\text{t}\nfw}} 
\newcommand{\dm}{{\textsc{dm}}}
\newcommand{\FD}{{\textsc{fd}}}

\def\mao{Main Astronomical Observatory, Nat. Acad. of Sci. of Ukraine}

\begin{document}
\title[New model of density distribution for fermionic dark matter halos]
{New model of density distribution for fermionic dark matter halos}%
\author{A.V.~Rudakovskyi}
\affiliation{\bitp}
\address{\bitpaddr}
\email{rudakovskyi@bitp.kiev.ua}
\affiliation{\knu}
\affiliation{\mao}
\author{D.O.~Savchenko}
\affiliation{\bitp}
\affiliation{\mao}

\udk{524.8} \razd{\seci}

\autorcol{D.O.\hspace*{0.7mm}Savchenko, A.V.\hspace*{0.7mm}Rudakovskyi}%

\setcounter{page}{1}%

\begin{abstract}
In this paper, we formulate a new model of density distribution for halos made of warm dark matter (WDM) particles.  The model is described by a single microphysics parameter\,---\,the mass (or, equivalently, the maximal value of the initial phase-space density distribution) of dark matter particles. Given the WDM particle mass and the parameters of a dark matter density profile at the halo periphery, this model predicts the inner density profile. In case of initial Fermi--Dirac distribution, we successfully reproduce cored dark matter profiles from $N$-body simulations. Also, we calculate the core radii of warm dark matter halos of dwarf spheroidal galaxies for particle masses $m_\FD= 100$, $200$, $300$ and $400$~eV.
\end{abstract}


\maketitle

The nature of dark matter\,---\,the largest gravitating substance in the Universe\,---\,is not yet identified.
Usual (left-handed) neutrinos\,---\,the only natural dark matter candidate within the Standard Model of particle physics\,---\,are too light to form the observed large-scale structure of the Universe~\cite{White:83} and the densest dark matter-dominated objects, dwarf spheroidals (dSphs)~\cite{Tremaine:79}. So far, many extensions of the Standard Model containing a viable dark matter candidate have been proposed; see, e.g., reviews~\cite{Bergstrom:00,Bertone:04,Feng:10,Gardner:13}. In terms of their initial velocities, valid dark matter candidates can be split in two groups\footnote{Note that, for some specific dark matter particle candidates, their initial velocity spectrum can be approximated by a mixture of `cold' and `warm' components \cite{Palazzo:07,Boyarsky:09a}.} (see, e.g., \cite{Primack:97}): 
\begin{itemize}
\item cold dark matter (CDM), composed of  particles with small (non-relativistic) initial velocities \cite{White:78,Blumenthal:84};
\item warm dark matter (WDM), composed of particles with large (relativistic) initial velocities \cite{Bisnovatyi-Kogan:80,Bond:80}. 
\end{itemize}

Density distribution of CDM haloes is often described by the Navarro--Frenk--White (NFW) profile \cite{Navarro:95,Navarro:96}
\eq{\rho_\nfw(r) = \frac{\rho_s r_s}{r \left(1+ \frac{r}{r_s}\right)^2}.\label{eq:rho-nfw}} Its parameters $\rho_s$ and $r_s$ are connected with the halo mass $M_{200}$ (the mass within the sphere of radius $R_{200}$, within which the average density is 200 times larger than the critical density $\rho_\text{crit}$ of the Universe) and halo concentration parameter $c_{200} = R_{200}/r_s$.

The phase-space density for CDM haloes becomes \emph{infinite\/} towards the halo centre; see, e.g., \cite{Taylor:01}. For WDM, this is not true: its maximal phase-space density $f_\text{max}$ is finite at early times and does not increase during halo formation \cite{Boyarsky:08a}. Usually, density distributions with finite $f_\text{max}$ are derived either from analytical studies of self-gravitating Fermi--Dirac dark matter (see, e.g., \cite{Ruffini:83,Bilic:97,Angus:09,deVega:13,deVega:14,Merafina:14,Domcke:14,Ruffini:14,Chavanis:14b,Arguelles:16}) or from $N$-body simulations imitating initial dark matter velocities (see, e.g., \cite{Shao:12,Maccio:12a,Maccio:12err,Maccio:12b,Anderhalden:12}).

The first method requires non-trivial assumptions about dark matter microphysics. Also, it often has problems with a \emph{simultaneous\/} description of the whole dark matter halo including the central part (where the dark matter phase-space density is close to $f_\text{max}$) and the outskirts (where it is $\ll f_\text{max}$). This, in turn, is well-established by $N$-body simulations. But simulations are computationally expensive; to determine the dark matter properties, one requires too many of them to compare to specific observations.

In this paper, we present a new model of density distribution that overcomes both difficulties. It is constructed in Sec.~\ref{sec:method} avoiding any assumptions about dark matter microphysics apart from the knowledge of the maximal value of dark matter phase-space density. As we show in Sec.~\ref{sec:comparison-Nbody}, this model predicts flattening of the inner density profile at small radii (producing dark matter \emph{cores\/}) consistent with WDM simulations \cite{Shao:12,Maccio:12a}.
 In Sec.~\ref{sec:core-radii}, we study the formation of dark matter cores for `classical' and `ultra-faint' dSphs.\footnote{For the dSphs nomenclature, see, e.g., \cite[Sec.~1.1]{Bullock:17}.} Finally, in Sec.~\ref{sec:discussion}, we discuss the obtained results.

\section{Method}\label{sec:method}
According to the strong Jeans' theorem~\cite{Jeans:1915,Lynden-Bell:62,Efthymiopoulos:07}, the phase-space density distribution $f(\vec{r},\vec{v})$ of a collisionless system in steady state depends on coordinates $\vec{r}$ and velocities $\vec{v}$ only through isolating~\cite{Contopoulos:63}
integrals of motion. Assuming a steady-state dark matter halo to be non-rotating, isotropic and spherically symmetric,\footnote{For observed dSphs, some of these assumptions may be violated. For example, Sagitarrius \cite{Ibata:94,Majewski:03}, Ursa Major~II \cite{Simon:07,Smith:13} and Bo\"{o}tes~III \cite{Carlin:09} dSphs are reported to be tidally disrupted, which puts the assumption of steady state in these objects under question. Basing on papers \cite{Kazantzidis:03,Hansen:05b,Zait:07,Sparre:12,Mamon:12,BeraldoeSilva:13,Vera-Ciro:14,El-Badry:16,Eilersen:17}, we expect only slight deviations from the velocity isotropy in the central parts of dark matter haloes. Although the authors of \cite{Hayashi:12,Hayashi:15} report deviations from spherical symmetry in dark matter haloes of dSphs, it is unclear to what extent their result can affect the density distribution. For example, according to \cite{Laporte:13}, the absence of spherical symmetry does not change the conclusion of \cite{Walker:11} about the presence of dark matter cores in dSphs (see, however, \cite{Genina:17}). Also, \cite{Campbell:16} shows that smaller galaxies tend to be more spherically symmetric and that dark matter distribution is more spherically symmetric than stellar distribution.
Finally, although rotations are detected in several individual objects (see, e.g., \cite{Simon:07,Amorisco:12a,Ho:12,delPino:16}), spectroscopic observations \cite{Walker:05,Koch:06,Frinchaboy:12,McConnachie:12,Spencer:17} show the absence of rotations with velocities comparable with the observed velocity dispersions in dSphs.} the phase-space density $f(\vec{r},\vec{v})$ of dark matter particles with mass $m_\FD$ inside the halo depends \emph{only\/} on their total energy $\mathcal{E}$, $E \equiv \frac{\mathcal{E}}{m_\FD} = v^2/2 + \Phi(r)$ \cite{Efthymiopoulos:07}, where $\Phi(r)$ is the local gravitational potential. 
 \eq{\Phi(r) = - 4\pi G_N\int_{r}^{\infty}\frac{dx}{x^2}\int_0^x \rho(y)y^2 dy,\label{eq:phi}} 
Under this assumption, the Eddington transformation~\cite{Eddington:1916,Binney-Tremaine:08book,Efthymiopoulos:07} unambiguously determines the phase-space density distribution given the dark matter density $\rho$:
\eq{f(E) = \frac{1}{\pi^2\sqrt{8}}\frac{d}{dE}\int_{E}^{0} \frac{d\rho}{d\Phi}\frac{d\Phi}{\sqrt{E-\Phi}}.\label{eq:fE}} 

We start from dark matter haloes with the NFW dark matter density distribution; see Eq.~(\ref{eq:rho-nfw}).
For such haloes, the phase-space density $f_\nfw(E)$ becomes infinite as $E\to \Phi(0) \equiv -4\pi G_N\rho_s r_s^2$~\cite{Widrow:00}.  This behaviour contradicts the expectations of the WDM model: according to the Liouville theorem, $f(E)$ should not exceed some finite maximal value $f_\text{max}$ of the initial phase-space density defined by dark matter microphysics. 
A particular example of interest is dark matter with initial Fermi--Dirac distribution having particle mass $m_\FD$ and $g$ internal degrees of freedom. For this dark matter model, the maximal value of the initial phase-space density is~\cite{Boyarsky:08a}
\begin{align}
&f_{\text{max}} = \frac{g m_\FD^4}{2(2\pi\hbar)^3} \nonumber \\ &= 1.31\times 10^4\left(\frac{g}{2}\right)\left(\frac{m_\FD}{400~\unit{eV}}\right)^4 M_\odot /\, \unit{kpc^{3}  (km/s)^{3}}\label{eq:fmax}
\end{align}
 (henceforth, we assume $g = 2$).
For any other dark matter particle model with known $f_\text{max}$, one can express it in terms of $m_\FD$ by using Eq.~(\ref{eq:fmax}).

To account for the maximal phase-space density, we \emph{truncate\/} $f(E)$ in a way that it cannot exceed the pre-selected maximal value $f_\text{max}$:
\begin{equation}
  \label{eq:19}
    f_\tnfw(E)\, {=}\, \left \{ 
    \begin{array}{ll}
     \!\!f_\nfw(E) , &   f_\nfw(E) < f_\text{max} \, ,\\
         \!\! f_\text{max}, & f_\nfw(E) \geq f_\text{max} \, .
        \end{array} \right. 
\end{equation}
The obtained phase-space density $f_\tnfw(E)$ is then converted to mass density via \cite{Binney-Tremaine:08book}
\eq{\rho_\tnfw(r) = 4\pi\int_{\Phi(r)}^0 f_\tnfw(E)\sqrt{2\left(E - \Phi(r)\right)} dE.\label{eq:rho-tnfw}}
Because, in Eq.~(\ref{eq:rho-tnfw}), the potential $\Phi(r)$ depends on the actual $\rho(r)$, we solve the system of equations (\ref{eq:fE})--(\ref{eq:rho-tnfw}) \emph{iteratively}.
We use the following iterative procedure:  we  calculate numerically the $\Phi_{i-1}(r)$ and $f_{i-1}(E)$  from the density distribution $\rho_{i-1}(r)$ obtained in the previous step.  Then we truncate $f_{i-1}(E)$ as in Eq.(\ref{eq:fE}) and obtain the new density distribution $\rho_i(r)$ from this truncated distribution function by using Eq.(\ref{eq:rho-tnfw}). We perform all the calculations on the grid in range $(r_0, r_\text{max})$. We choose $r_0\ll r_s$ for regularization at the first iteration. The  $r_\text{max}$ is defined as $\rho_\nfw(r_\text{max})={\bar\rho_\dm}$ and $r_\text{max} \gg R_{200}$, so we use it as the upper limit of integration in Eq.~(\ref{eq:phi}). We use the value $\text{max}\left|\frac{\rho_i(r)-\rho_{i-1}(r)}{\rho_{i-1}(r)}\right|$ as convergence criterion. As demonstrated in Fig.~\ref{fig:rho-our-vs}, five iterations is sufficient to achieve convergence for our chosen grid parameters ($\text{max}\left|\frac{\rho_5(r)-\rho_4(r)}{\rho_4(r)}\right|<0.01$). We demonstrate the convergence of this iterative procedure by the numerous numerical tests and do not strictly prove it.  The obtained results show the very weak dependence of obtained truncated density profiles on the grid parameters.

\section{Results}
\label{sec:results}

\subsection{Comparison with $N$-body simulations}\label{sec:comparison-Nbody}

To check the validity of the proposed approach, we compared the truncated dark matter density distribution $\rho_\tnfw(r)$ to the results of two independent $N$-body simulations \cite{Shao:12,Maccio:12a}.
Both simulations include the effect of maximal phase-space density by assigning non-zero initial velocities to dark matter particles. More precisely, \cite{Shao:12} assumes the Fermi--Dirac distribution $f(v) = \left[\exp(v/v_0)+1\right]^{-1}$, where $v_0$ is the characteristic velocity of dark matter particles \cite{Bode:00}, while \cite{Maccio:12a} approximates it with a \emph{Gaussian\/} velocity distribution. 

\begin{figure*}
\vskip1mm
\includegraphics[width=\column]{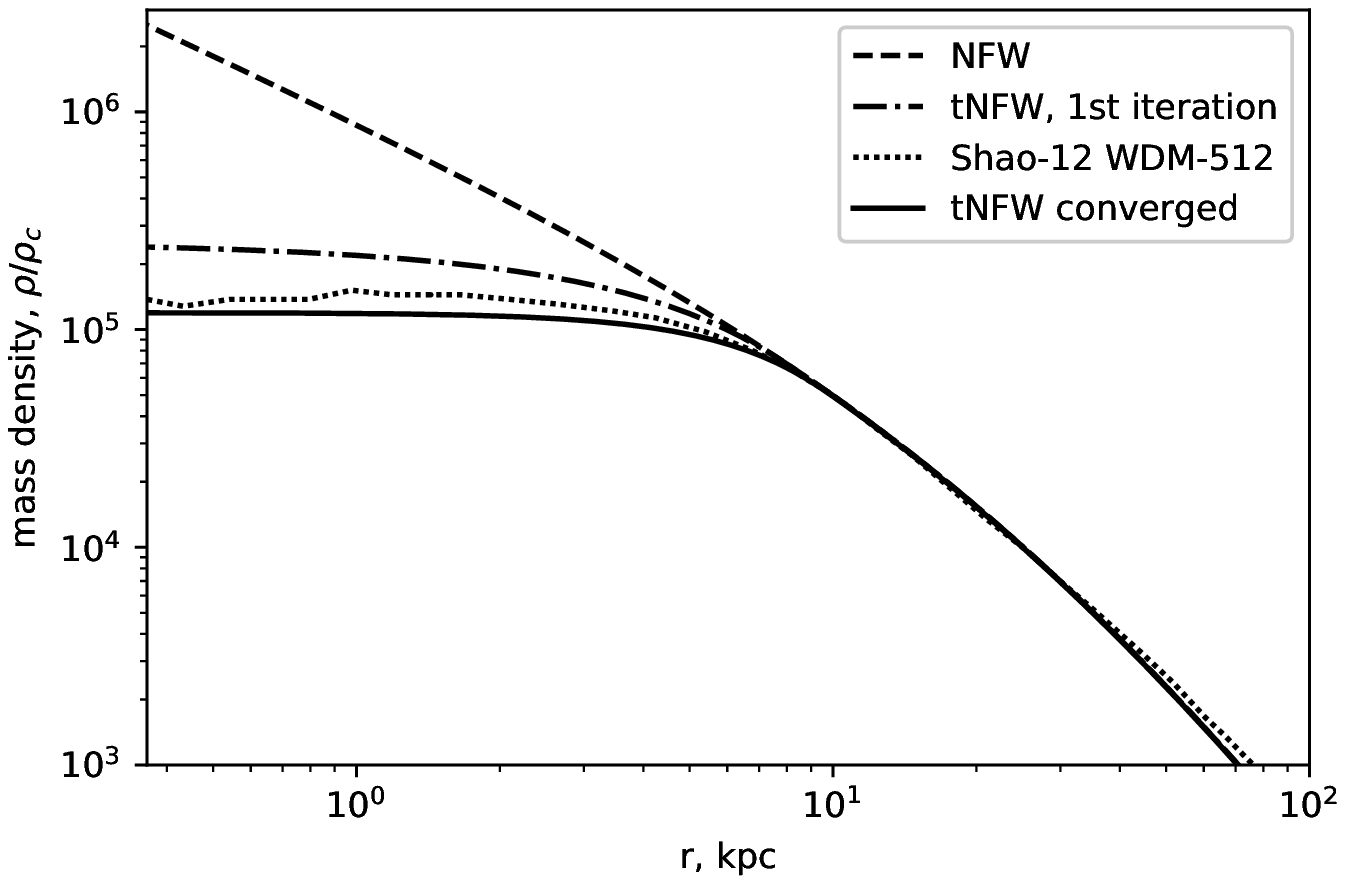}
\includegraphics[width=\column]{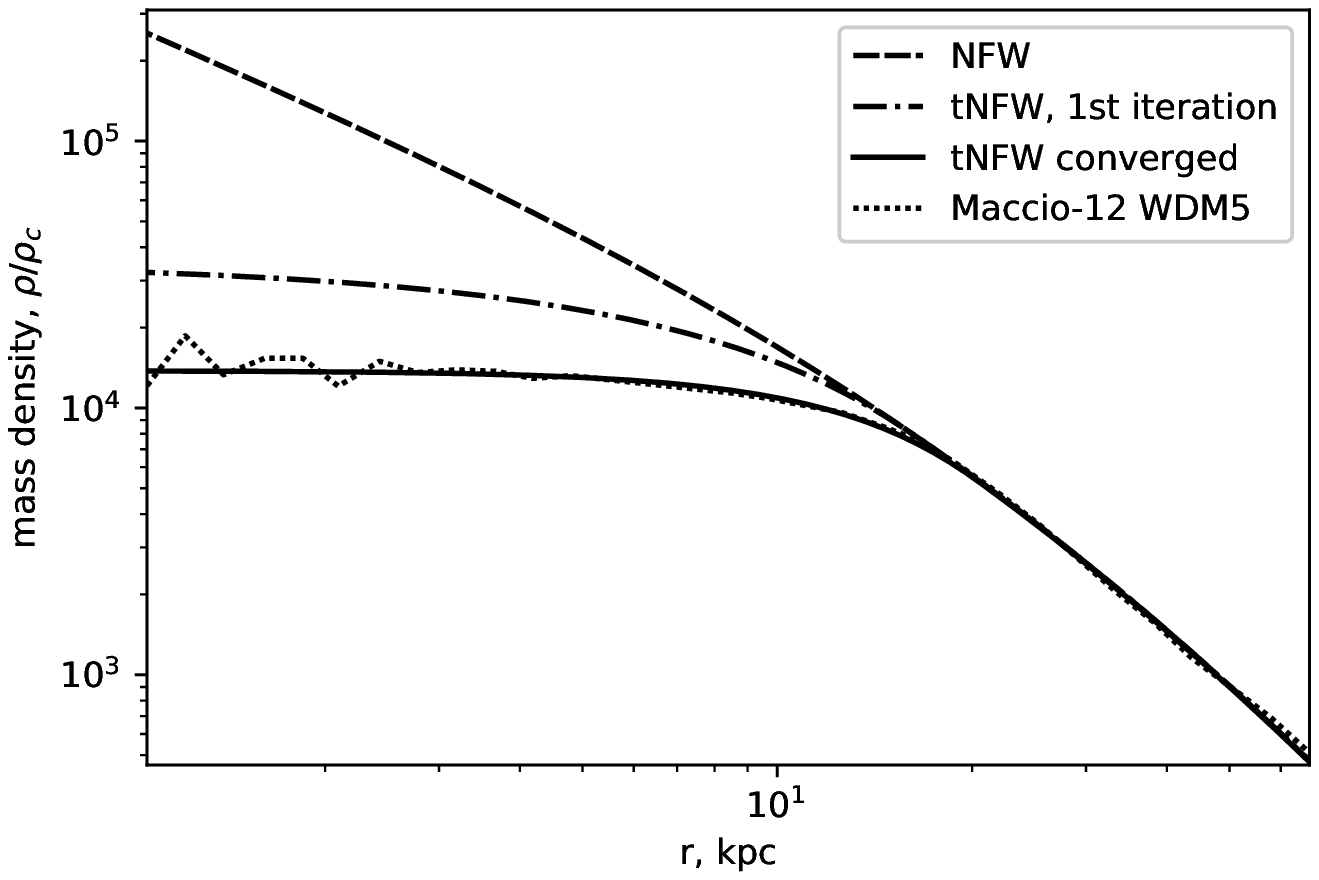}
\vskip-3mm
\caption{\label{fig:rho-our-vs} \textit{Left\/}: Comparison of tNFW profile with $N$-body simulation P-WDM$_{512}$ of \cite{Shao:12}.
We started from the NFW profile with $\rho_s = 4.9\times 10^{7}~\unit{M_\odot/kpc^3}$ and $r_s = 26~\unit{kpc}$ corresponding to the 
parameters of P-WDM$_{512}$ halo in Fig.~2 of~\cite{Shao:12}. Then, we built the tNFW profile with $m_\FD = 30$~eV using equations (\ref{eq:fE})--(\ref{eq:rho-tnfw}). For individual iterations, mass density profiles (shown by black dash-dotted curve) started to converge quickly, so we used the sixth iteration (shown by a solid curve) as a final tNFW profile. Due to truncation of phase-space density distribution, the final value of $M_{200}$ for this tNFW profile becomes $1.27\times 10^{12}~\unit{M_\odot}$, decreasing from its initial value by 9\%. Note the similarity between the tNFW profile and the result of $N$-body simulations of \cite{Shao:12} (dotted curve): the resulting mass densities and core radii differ by less than 30\%. \textit{Right\/}: comparison of the density distribution from simulation WDM-5 of \cite{Maccio:12a} and the corresponding tNFW profile starting from $\rho_s = 1.2\times 10^{6}~\unit{M_\odot/kpc^3}$, $r_s = 33~\unit{kpc}$  and using $m_\FD = 23$~eV (see text). For this tNFW profile, the final value of $M_{200}$ is $4.7\times 10^{11}~\unit{M_\odot}$, decreasing from the initial value by 11\%. The differences between the density profiles and derived core radii are also $\lesssim 30\%$ in this case.}
\end{figure*}

In Fig.~\ref{fig:rho-our-vs}, we compare the tNFW density profile to simulation P-WDM$_{512}$ from~\cite{Shao:12} and simulation WDM-5 from~\cite{Maccio:12a}, corresponding to $m_\FD = 30$~eV and 23~eV, respectively.\footnote{The last number differs from the value reported in Table~1 of \cite{Maccio:12a} To obtain it, we took into account that \cite{Maccio:12a} assumed Gaussian velocity distribution with velocity dispersion $\sigma = 3.571 v_0$ \cite{Bode:00}, so that their maximal phase-space density can be calculated from the dark matter density $\rho_\dm$: $$f_\text{max} = \frac{\rho_\dm}{\left(2\pi\sigma^2\right)^{3/2}}.$$ For WDM-5 simulation of~\cite{Maccio:12a},
$f_\text{max} = 0.151~M_\odot /\,\unit{kpc^3 (km/s)^3}$, which corresponds to $m_\FD = 23$~eV according to  Eq.~(\ref{eq:fmax}).} We extracted the NFW parameters from Fig.~2 of \cite{Shao:12} and Fig.~2 of \cite{Maccio:12a}, respectively, and calculated the corresponding tNFW profiles. Fig.~\ref{fig:rho-our-vs} shows that  tNFW profiles match the corresponding WDM distributions at the $\lesssim 30\%$ level. Also, we do not observe any systematic disagreement between the tNFW profile and other WDM profiles from $N$-body simulations~\cite{Shao:12,Maccio:12a,Maccio:12b}.

\subsection{Core radii of dwarf spheroidal galaxies}\label{sec:core-radii}

An important property of the distribution $\rho_\tnfw(r)$ is its flattening towards small radii. This flattening is usually characterised by the radius of the so-called \emph{dark matter core\/}.\footnote{There are many different definitions of core radii in the literature; see, e.g., \cite{Randall:16,Maccio:12a}.} 
In this paper, we define core radius $r_c$ for a given dark matter distribution $\rho_\dm(r)$ as follows: \eq{\rho_\tnfw(r_c) = \frac{\rho_\tnfw(0)}{4}.} This definition coincides with the characteristic radius of the widely used Burkert density distribution~\cite{Burkert:95}, as well as with the core radius defined in \cite{Domcke:14,DiPaolo:17}. The largest effect from finite phase-space density on the core sizes is expected in the systems hosted by the densest dark matter haloes\,---\,dwarf spheroidals (dSphs).

We analysed the two types of halos of dwarf spheroidals: `classical' dSph with corresponding NFW parameters  $M_{200} = 4 \times 10^8~\unit{M_\odot}$ and $c_{200} = 30$, and `ultra-faint' dSph with $M_{200} = 1 \times 10^8~\unit{M_\odot}$ and $c_{200} = 40$. Assuming initial Fermi--Dirac distribution of warm dark matter particles, we generated tNFW profiles with $m_\FD = 100$, 200, 300 and 400~eV for these halos. The obtained core radii $r_c$ are summarised in Table~\ref{table:1}.

\begin{table}[t]
\vskip4mm
\noindent\caption{\label{table:1}Core radii for tNFW density profiles of `classical' and `ultra-faint' dwarf spheroidal galaxies for $m_\FD = 100$, $200$, $300$ and $400~\unit{eV}$.}\vskip3mm\tabcolsep4.5pt   

\noindent{\footnotesize
\begin{tabular}{|c|c|c|}
\hline
    \rule{0pt}{5mm}dSph & $m_\FD$, eV& $r_c$,~ kpc\\
    \hline
   \rule{0pt}{5mm}`classical' &100&3.82  \\
    
    &200&0.86  \\
    
    &300&0.41  \\
    
    &400&0.26  \\
    \hline
\rule{0pt}{5mm}`ultra-faint' & 100&5.57\\
    &200& 1.08 \\
    &300& 0.46 \\
    &400&0.27  \\
\hline
\end{tabular}
}
\end{table}

\section{Discussion}
\label{sec:discussion}

In this paper, we described a new simple model to quantify the effect of maximal phase-space density ($f_\text{max}$) on dark matter distribution. In contrast to other models, we started from the well-known Navarro--Frenk--White (NFW) density profile~\cite{Navarro:95,Navarro:96} consistent with $N$-body simulations with $f_\text{max} = \infty$.  After that, we truncated the corresponding phase-space density profile at the value of $f_\text{max}$ and recalculated the corresponding density profile using the Eddington transformation \cite{Eddington:1916}. The obtained `truncated NFW' (tNFW) density distribution flattens at small radii producing a \emph{core\/}. Despite its simplicity, tNFW profile matches the detailed $N$-body simulations from \cite{Shao:12,Maccio:12a} with high precision, $\lesssim 30\%$ as demonstrated in Fig.~\ref{fig:rho-our-vs}.

Recent papers \cite{Domcke:14,DiPaolo:17} have questioned the lower bound on $m_\FD \gtrsim 0.48$~keV obtained in \cite{Boyarsky:08a}. By assuming anisotropic velocity distribution, the authors of \cite{Domcke:14} claimed that velocity dispersion profiles in `classical' dSphs are consistent with the mass of degenerate fermions as low as $m = 0.2$~keV (equivalent to $m_\FD = 0.24$~keV). Paper \cite{DiPaolo:17} extends this result by using data on both `classical' and `ultra-faint' dSphs. According to~\cite{DiPaolo:17}, the lower bound of \cite{Boyarsky:08a} should be further weakened down to $m_\FD \gtrsim 0.12$~keV in the case of arbitrary stellar velocity anisotropy and no relation between stellar and dark matter velocity dispersions.

It is suggested in \cite{Amorisco:12b} that the Fornax dSph has a core with $r_c= 1_{-0.4}^{+0.8}$~kpc. 
To create sizeable cores in halos of dwarf galaxies without involving baryonic processes, one requires rather light dark matter particles ($m_\FD<300$~eV; see Table~\ref{table:1}). Such dark matter is in tension with a number of constraints from observations of large-scale structure; see, e.g., recent works \cite{Read:16,Corasaniti:16,Schneider:16,Menci:17a,Cherry:17,Birrer:17,Irsic:17,Yeche:17,Lopez-Honorez:17,Dayal:17,Baur:17,Menci:17b}. However, due to the smallness of their potential wells, dwarf galaxies are very sensitive to baryonic feedback processes; see, e.g., \cite{Dekel:86,Ferrara:99,Read:04,Read:06,Mashchenko:07,Pontzen:11,Governato:12,Teyssier:12,DiCintio:13,Read:15}. Working together with the finite dark matter phase-space density effect, these processes could produce much larger cores, with core sizes close to $\sim 1$~kpc.

\textit{
The authors are grateful to D.~Iakubovskyi and Yu.~Shtanov for collaboration and valuable comments. This work was partially supported by the National Academy of Sciences of Ukraine (project No. 0116U003191), by the Program of Fundamental
Research of the Department of Physics and Astronomy of the National
Academy of Sciences of Ukraine (project No. 0117U000240), by grant 6F of the Department of Targeted Training of the Taras Shevchenko Kyiv National University under the National Academy of Sciences of Ukraine, and by the
Program of Cosmic Research of the National Academy of Sciences of Ukraine.}

\bibliography{preamble_ujp,PSD-bound-kinematics}

\end{document}